\documentclass[5p]{elsarticle}

\usepackage{graphicx}% Include figure files
\usepackage{dcolumn}% Align table columns on decimal point
\usepackage{bm}% bold math
\usepackage{longtable}
\usepackage{float}
\usepackage[fleqn]{amsmath}
\usepackage{amsfonts}

\makeatletter
\def\ps@pprintTitle{%
 \let\@oddhead\@empty
 \let\@evenhead\@empty
 \def\@oddfoot{}%
 \let\@evenfoot\@oddfoot}
\makeatother

\usepackage{lineno,hyperref}
\modulolinenumbers[5]

\begin{document}

\begin{frontmatter}

\title{Heat diffusion in magnetic superlattices on glass substrates}

\author{F. Hoveyda \fnref{fn1}}

\author{M. Adnani \fnref{fn1,fn2}}

\author{S. Smadici \corref{cor1}}

\fntext[fn1]{Equal first author}

\fntext[fn2]{Present address: Department of Physics, University of Houston, Houston, TX 77204, USA}

\address{Department of Physics and Astronomy, University of Louisville, KY 40292, USA}%

\begin{abstract}
Pump-probe experiments and polarizing microscopy are applied to examine temperature and heat flow in metallic magnetic superlattices on glass substrates. A model of heat diffusion in thin layers for cylindrical symmetry, equivalent to the Green's function method, gives a good description of the results. The frequency dependence of temperature modulation shows that a glass layer should be added to the sample structure. The demagnetization patterns are reproduced with a Green's function that includes an interface conductance.
\end{abstract}

%\begin{keyword}
%Heat diffusion; heterostructure; ultrafast laser; polarizing microscopy
%\end{keyword}

\end{frontmatter}

%\linenumbers

%\maketitle

\section{Introduction}

Thermal energy management of opto- and spintronic devices under pulsed laser excitation becomes increasingly important as devices decrease in size. Heat diffusion in multilayer materials can be quantified with modulated thermoreflectance measurements of the temperature spatial profile~\cite{2007Fretigny} and frequency dependence~\cite{1994Reichling}. Pump-probe time-domain measurements of the temperature time-evolution, following a TiS laser ultrafast transient disturbance $T_{tr}$, developed into a powerful technique for measuring thermal conductivity and interface conductance~\cite{2014Cahill}. An additional heat accumulation temperature increase $T_{acc}$, similar to that of modulated thermoreflectance, will also result for these sources, when the thermal energy deposited into a highly-absorbing metal layer does not fully dissipate between pulses~\cite{2008Schmidt-a,2008Schmidt-b}.

A magnetic material saturation magnetization $M_{s}$, magnetic anisotropy $K$, exchange energy $A$, coercive field $H_{c}$ all depend on $T$. The equilibrium magnetization magnitude and direction is determined by a balance of several energies that depend on these factors and implicitly on temperature. For instance, temperature increases induced by laser beams can modify the equilibrium conditions and start a magnetization precession, switch it between different easy axis minima~\cite{2017deJong,2017Afanasiev}, or modify the magnetization hysteresis loops in applied fields~\cite{2007Stanciu}. Magnetic structures induced by laser beams~\cite{2006Aktag,2006Schuppler,2009Leufke,2014Kisielewski,2015Stark}, all-optical switching (AOS) in thin ferrimagnetic rare-earth and ferromagnetic superlattices~\cite{2014Lambert,2014Mangin}, hybrid structures~\cite{2017Gorchon} and granular media~\cite{2016Takahashi,2017John} have been observed. Local ultrafast~\cite{2007Stanciu,2016Cornelissen} as well as thermal models of AOS~\cite{2012Khorsand,2012Ostler,2016Xu,2016Ellis,2016Gorchon} have been proposed. Recently, cumulative AOS in ferrimagnetic rare earth/transition metal, ferromagnetic Co/Pt~\cite{2016Hadri} and Co/Pd~\cite{2017F} superlattices showed how thermally-induced forces can move magnetic domain walls. For a quantitative understanding of these observations it is necessary to characterize the thermal response of the samples.

In this work, metallic magnetic superlattices were examined with transmission pump-probe measurements of modulated heat accumulation $T_{acc}$ and polarizing microscopy. Two-dimensional heat diffusion and thermal demagnetization patterns illustrate the energy flow in the structures. Green's functions are calculated for different thermal properties and sample geometries in the two experimental configurations of a chopped and a moving beam.

\section{Experimental setups}

\subsection{Samples}

$\rm [Co/Ag]_{3}$, $\rm [Co/Pd]_{4}$ and $\rm [Co/Au]_{10}$ multilayer samples were deposited with e-beam evaporation on 1 mm thick soda-lime glass substrates at room temperature. For $\rm Co/Ag$, the rates were adjusted to give 1 nm Co and 1.4 nm Ag individual layer thicknesses and the sample was capped in-situ with a $\rm SiO_2$ layer. The $\rm Co/Pd$ and $\rm Co/Au$ glass substrates were immersed in Nanostrip solution for five minutes, then placed in acetone and methanol, and sonicated in each liquid for 10 minutes at $60^0$ C. Hot Nanostrip ($90^0$ C) was used for cleaning the $\rm Co/Au$ glass substrates. During the deposition, the substrates were rotated at $5$ RPM about an axis making an angle of $45$ degrees with the surface normal. Samples of variable thickness were also made with a custom rotating holder that obscured the source over a position-dependent fraction of the deposition time.

Profilometry measurements on Co/Ag showed a maximum total sample thickness of $\rm 15.6~nm$. The thickness variation along the sample surface was obtained with transmission and reflection measurements with a 633 nm He-Ne laser. For instance, for Co/Ag, the thickness decreased linearly along the surface up to 28 mm, beyond which it drops rapidly (figure 1(b), inset). The $\rm [Co/Pd]_{4}$ and $\rm [Co/Au]_{10}$ sample thickness were 6.2 nm and 49 nm.

\subsection{Pump-probe measurements}

A two-frequency pump-probe setup was applied in a non-collinear transmission geometry. The pump and probe beams were focused on the sample to $w_{0}=125~\mu m$ and $w_{1}=80~\mu m$ spots, respectively. The linearly-polarized pump beam of $120~fs$ pulses at a repetition rate of $80~MHz$ and $800~nm$ wavelength was chopped and the polarization variation of a low-fluence, linearly-polarized, 400 nm probe beam was measured with a Wollaston prism, balanced photodiode and lock-in amplifier. The probe beam was aligned along the surface normal. The incidence angle of the pump beam was $10$ degrees relative to the surface normal. A stage varied the delay between the two pulse sequences (figure 1(a)).

Heat accumulation occurs when the interval between consecutive pulses $\frac{1}{f_{rep}}=12.5~ns$ is smaller than the time it takes heat to diffuse out of the beam footprint $\frac{w_{0}^{2}}{4D} = \frac{(125~\mu m)^{2}}{0.4~cm^{2}/s} = 0.4~ms$ with typical metal diffusivity. The temperature $T(t)=T_{tr}+T_{acc}$ is made of two components: a transient $T_{tr}$ peak, related to ultrafast non-thermal processes, and a heat accumulation $T_{acc}$ from the cumulative effect of multiple pulses, as illustrated by a time-resolved measurement (figure 1(a), inset). $T_{step}$ is the small temperature increase due to one pulse that dissipates before the next pulse arrives and, as expected from the estimate above, is only a small fraction of $T_{acc}$.

\begin{figure}
\centering\rotatebox{0}{\includegraphics[scale=0.35]{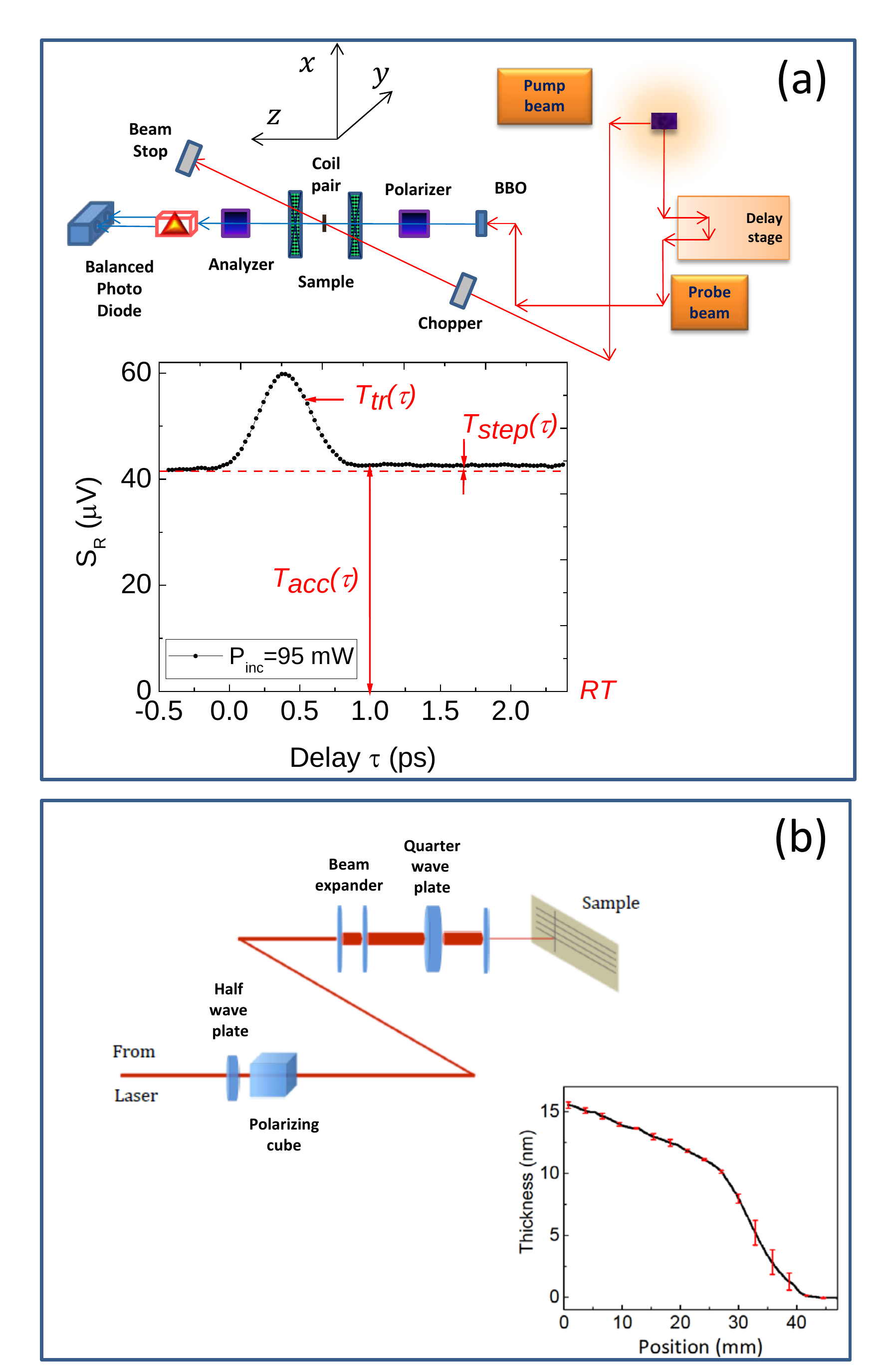}}
\caption{\label{fig:Figure1} (a) Sketch of the pump-probe setup. Inset: the response of the sample symmetric in applied field $S(\tau)=\frac{1}{2}\Big( S(+B)+S(-B) \Big)$ measured at $f_{ch}=2.069~kHz$. (b) The scanning setup with the Co/Ag sample thickness profile in the inset.}
\end{figure}

The pump $F_{0}(\omega)$ and probe $F_{1}(\omega)$ beam fluences have the spectrum of the femtosecond comb. In thermoreflectance experiments a sinusoidal intensity modulation of large frequency ($> 0.1~MHz$) is usually applied~\cite{2004Cahill}. In our experiments, we vary the pump beam fluence with a square-wave chopper modulation. This introduces additional sidebands in the spectrum at $nf_{ch}$, where $n$ is an integer. However, the lock-in detects at the chopper modulation frequency $f_{ch}$. This is equivalent to replacing the pump pulse sequence with $F_{0}(\omega) \rightarrow F_{0}(\omega_{ch})$, keeping only the temporal profile at the frequency $f_{ch}=\frac{\omega_{ch}}{2 \pi}$, and the probe pulse sequence with the average fluence $F_{1}(t) \rightarrow const$ ~\cite{2009Schmidt}. The pump fluence modulation gives a  relatively slowly-varying $T_{acc}$ oscillation which, as expected, does not depend on delay within a few tens of $ps$, either before or after the overlap.

An analysis similar to that done in $\omega, k$ variables for time-resolved pump-probe measurements can be applied~\cite{2004Cahill}. Specifically, the absorbed fraction of the incident pump pulse fluence $F_{0}(\omega,k)$ is converted to an initial temperature $T_{initial}(\omega,k)=\frac{F_{0,abs}(\omega,k)}{Cd}$ with the area specific heat $Cd$, where $C$ is the volume specific heat. This temperature evolves to a final temperature $T(\omega,k)$ with the Green's function $\mathbb{G}(\omega,k)$ representing the heat diffusion. The final temperature $T(\omega,k)$ is sampled by the probe pulse $F_{1}(\omega,k)$ into  a spatially-averaged temperature $T(\omega)$.

We apply a Faraday Effect transmission geometry for our semi-transparent samples, complementary to the Kerr Effect reflection geometry on magnetic layers~\cite{2014Liu} and detect intensity and polarization variations. Then, the lock-in amplifier signal is

\begin{eqnarray}
L(\omega_{ch}) = A \mathbb{T'}(T(\omega_{ch})) =B T(\omega_{ch}) \\ \nonumber
=B \int dk k F_{1} (k) T(\omega_{ch},k)  \\ \nonumber
=B \int dk k F_{1}(k) \frac{F_{0,abs}(k)}{Cd} \mathbb{G}(\omega_{ch}, k)
\end{eqnarray}

\noindent where $F_{0,1}(k)$ are the Hankel transform of the two beams spatial profiles, the spatial averaging is done by combining the pump and probe beam profiles into a $dk$ integral~\cite{2004Cahill}, and $\mathbb{T'}$ is one of the several material properties that depend on temperature. The constant $A$ accounts for different units of $L$ (in volts) and $\mathbb{T'}$. The non-linear terms in $\mathbb{T'}(T)$ have been removed since they give a signal at multiples of $f_{ch}$ and $B\equiv A \frac{d\mathbb{T'}}{dT}$.

Part of the signal is proportional to the polarization rotation $\theta_{M}=-\frac{\pi d}{\lambda}nQ_{z}$ for a beam propagating along the $z-$axis due to a magnetization $M_{z} \propto Q_{z}$, where $Q_{z}$ is the off-diagonal magneto-optical coefficient in the susceptibility matrix~\cite{2000Qiu}. This part is anti-symmetric in $B$, can be removed in a combination $S=\frac{1}{2}\Big( S(+B)+S(-B) \Big)$, is relatively small~\cite{2017F-arXiv} and is neglected here. The part of the signal symmetric in $B$ is a thermal modulation of the transmittance $\mathbb{T}$, due to the temperature coefficient of refractive index $\frac{dn}{dT}$. The factor $\frac{d\mathbb{T}}{dT}$ may be called ``thermo-transmittance", by analogy with the complementary thermoreflectance $\frac{dR}{dT}$, which has been examined in detail for a series of materials~\cite{2010Wang} at different wavelengths~\cite{2012Wilson}. For metals $\frac{dn}{dT}=1-5\times 10^{-5}~K^{-1}$~\cite{2010Wang,2012Wilson}, significantly larger than $\frac{dn}{dT}=2\times 10^{-6}~K^{-1}$ for our ($\rm 17.5~\%~ Na_{2}O$, $\rm 7.5~\%~CaO$, $\rm 75~\% ~SiO_{2}$) soda-lime glass substrate~\cite{1993Jewell}. We then obtain a signal which is proportional to the temperature of the metallic film, where the $z$-dependence of temperature can be neglected inside our thin thermally-conducting samples.

\subsection{Scanning measurements}

The beam was expanded and focused on the front of the sample surface with a 30 mm lens to a typical $w_{0}=50~\mu m$ spot (figure 1(b)). The sample was scanned under the beam at a speed $v_{s}=10~\rm{mm/s}$ at constant fluence. Fluence was adjusted between scans with a half-wave plate and polarizing cube combination. Polarization was adjusted from linear to left- and right-circular polarized with a quarter wave plate. No variation was observed with changes in beam polarization.

Polarizing microscopy images were made in transmission Faraday geometry at normal incidence. In contrast to rotation $\theta_{M}$ from magnetization-induced birefringence, the rotation due to structural birefringence depends on the orientation of sample birefringence axes and can be varied with sample rotation.

New areas are continuously exposed in writing experiments with a moving beam. The diffusion time $w_{0}^2/4D = 0.4~ms$ out of the beam footprint is comparable to the moving beam dwell time $\tau_{l} = \frac{w_{0}}{v_{s}} = 5~ms$. The larger radius observed at the end of the stripe (figure 3(c)) confirms that the steady-state is not obtained during scanning.

\section{Results}

\begin{figure}
\centering\rotatebox{0}{\includegraphics[scale=0.2]{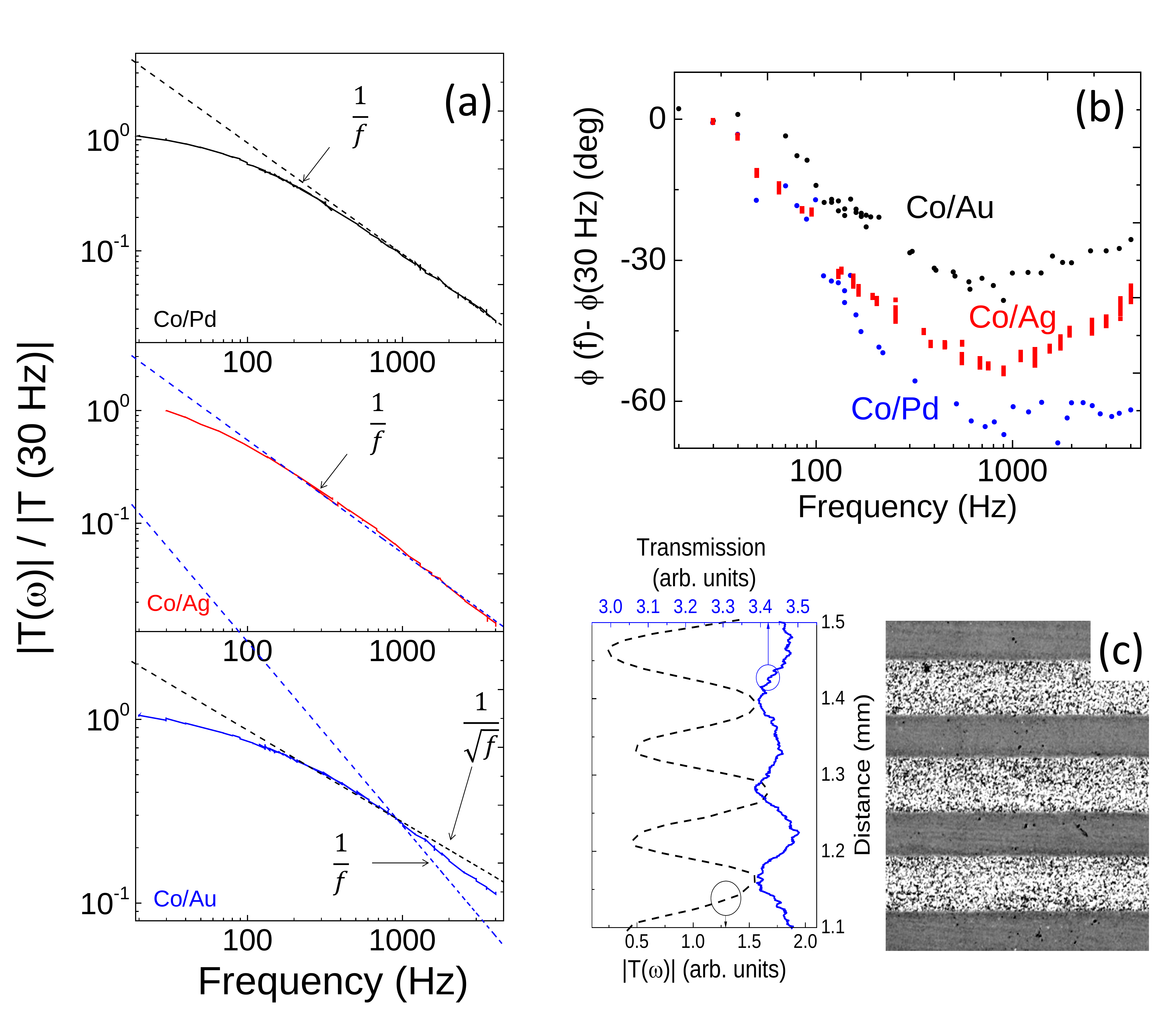}}
\caption{\label{fig:Figure1} (a) Temperature modulation amplitude $|T(\omega)|$ dependence on $f_{ch}=\frac{\omega_{ch}}{2 \pi}$, normalized to the value at 30 Hz, for Co/Pd ($40~mW$ incident power), Co/Ag ($46~mW$), and Co/Au ($40~mW$). Similar results were obtained for other powers between $20~mW$ and $60~mW$. (b) Phase dependence on chopper frequency $f_{ch}$. (c) Light transmission is higher at stripe locations and correlates with variations in $|T(\omega)|$ across Co/Ag stripes.}
\end{figure}

Measurements of temperature oscillation amplitude $|T(\omega)|$ and phase $\phi(\omega)$ dependence on the chopper frequency $f=\frac{\omega}{2 \pi}$ were made at a $\tau=-2~ps$ delay. Amplitude and phase are plotted relative to measurements at $\rm 30~Hz$. The amplitude decreases inversely proportional to $f$ above $200~Hz$ for Co/Pd and Co/Ag and in-between $1/f$ and $1/\sqrt{f}$ for Co/Au (figure 2(a)). A levelling of the amplitude is observed for frequencies below $200~Hz$. The phase $\phi(\omega)$ also varies with $f$, first decreasing and then slightly increasing, with a minimum at $f= 1000~Hz$.

Beam scans across the surface give ``stripes", with small white and black dots at stripe center corresponding to small magnetic domains oriented up and down and domain walls pinned by imperfections (figure 2(c)). Pump-probe measurements showed a reduced modulation amplitude $|T(\omega)|$ at stripe locations and light transmission measurements, made with a small intensity 800 nm TiS beam, confirmed that these locations have a smaller absorption $F_{0,abs}$ compared to pristine areas (figure 2(c)). A second stripe, made at constant fluence and sample thickness, gradually disappears when intersecting a stack of stripes (figure 3(a)) and re-appears intact once the stack is crossed.

Stripe edges are birefringent, as shown by intensity variations as the analyzer is rotated across the extinction condition [figure 3(b) and supplementary figure 1]. The birefringence is structural, since it varies from bright to dark over a 90 degrees sample rotation angle. The orientation of the sample birefringence axes depends on the orientation of light polarization when writing the stripes. As expected, birefringence is absent at the intersection of two stripes, made with light beams with orthogonal polarizations (not shown). A slight $30~nm$ bulging was also observed with AFM across stripes made at high power. Similar results were obtained for the other samples.

Isotropic and birefringent modifications can be made in clear glass with amplified TiS lasers following non-linear multi-photon absorption~\cite{2001Sudrie}. Cumulative heating has been considered for isotropic structural changes made with un-amplified lasers~\cite{2001Schaffer,2003Schaffer}. The higher fluence birefringence can arise from stress or $\mu m$-size elongated voids made in explosive processes of multiphoton and avalanche ionization~\cite{2005Hnatovsky,2009Cheng}. No signal is detected from clear glass for our relatively low fluence and negligible non-linear absorption. Therefore, because linear light absorption in the metal film is required, laser-induced changes in the glass substrate are made in the immediate vicinity of the metal film.

\begin{figure}
\centering\rotatebox{0}{\includegraphics[scale=0.35]{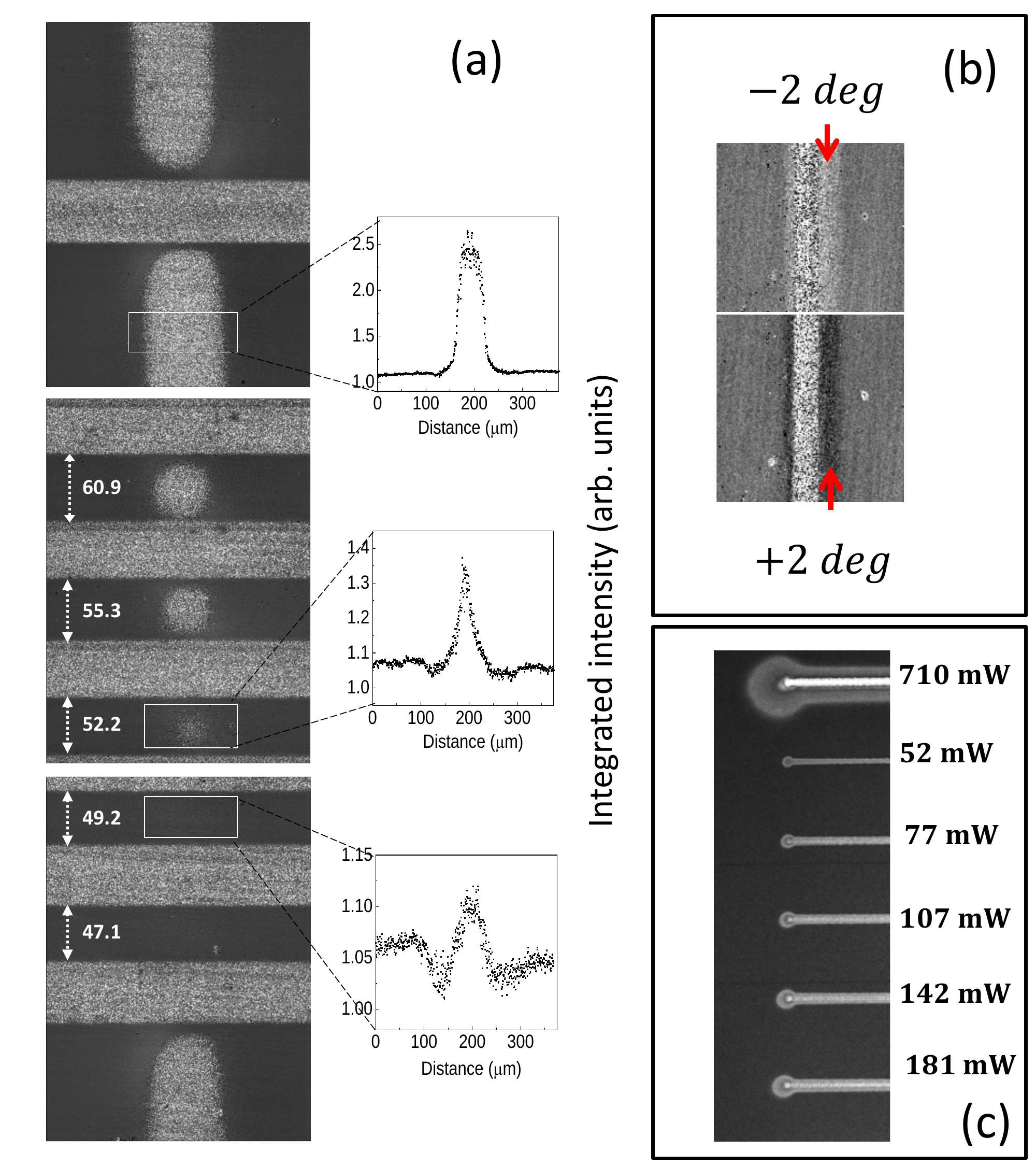}}
\caption{\label{fig:Figure1} (a) Vertical stripe made in Co/Ag across a stack of horizontal stripes, with distances shown in $\mu m$. Integrated intensity profiles show the stripe almost completely disappears when inside the stack, while the edges are relatively unaffected. (b) Structural birefringence at stripe edges. (c) End of stripe for Co/Au at different incident power shows the steady-state condition is not obtained during scanning.}
\end{figure}

\section{Discussion}

\subsection{Heat diffusion in multilayers}

Ultrafast processes occur within the first few $ps$, until equilibration to a common temperature $T=T_{e}=T_{latt}$. The subsequent time-evolution of $T$ in the structure is determined by the heat diffusion equation. It is advantageous to solve this equation following the method applied in time-resolved thermoreflectance in cylindrical symmetry, with new $(\omega, k, z)$ variables replacing $(t,r,z)$. The small ellipticity of the pump beam footprint is neglected. For instance, for the temperature, $k$ and $r$ are related as

\begin{eqnarray}
T(t,r,z)=\int_{0}^{\infty}dk k J_{0}(kr)\int_{-\infty}^{+\infty} e^{i\omega t}T(\omega,k,z)
\end{eqnarray}

\noindent where $J_{0}(x)$ is the Bessel function~\cite{2004Cahill}. The reverse $r \rightarrow k$ transform is $T(k)=\int_{0}^{\infty}dr r J_{0}(kr)T(r)$. Time $t$ and $\omega=2\pi f$ are related by the usual Fourier transform. The heat diffusion equation with no sources becomes in these variables $\frac{\partial ^{2}T(\omega,k,z)}{\partial z^{2}}=q^{2}T(\omega,k,z)$, where $q^{2}=\frac{\Lambda_{\parallel}k^{2}+i\omega C}{\Lambda_{\perp}}=k^{2}+\frac{i\omega}{D}$, $\Lambda_{||}=\Lambda_{\perp}$ are the thermal conductivities parallel and perpendicular to the surface, and $C$ is the volume specific heat. Its solutions are hyperbolic trigonometric functions that can be arranged in a matrix, describing how the temperature and flux vary with depth $z$.

Similar transformations are applied to all functions of $t$ and $r$, in particular to the beam profiles. The spatial dependence of a Gaussian beam fluence $F(t,r)=F_{peak}e^{-\frac{2r^{2}}{w_{0}^{2}}}=\frac{2P_{abs}}{\pi w_{0}^{2}}e^{-\frac{2r^{2}}{w_{0}^{2}}}$, where $F_{peak}$ has been replaced with the absorbed power $P_{abs}=\int_{0}^{\infty}dr 2\pi r F_{peak} e^{-2\frac{r^{2}}{w_{0}^{2}}}=\frac{\pi w_{0}^{2}F_{peak}}{2}$, has a Hankel transform $F(\omega,k)=\frac{P_{abs}}{2\pi}e^{-\frac{k^{2}w_{0}^{2}}{8}} $. The in-plane averaging of the pump-induced temperature by the probe is a pump-probe profile convolution in real space or a multiplication in $k-$ space~\cite{2004Cahill}, giving an effective diameter $w_{eff}=\sqrt{w_{0}^{2}+w_{1}^{2}}=150~\mu m$. Then, the characteristic magnitude of $k$ in our case is $k_{eff}=\frac{4}{150}~\mu m^{-1}\approx 0.025~\mu m^{-1}$.

The fluxes $F_{t,b}$ and temperatures $T_{t,b}$ on the top (front) and back sides of a multilayer in the limit of a thermally thin film ($q_{f}d_{f}=q_{f}d \ll 1$) and thermally thick substrate ($q_{s}d_{s} \gg 1$) are related by

\begin{eqnarray}
\Bigl(\begin{smallmatrix} T_{b}\\ F_{b} \end{smallmatrix} \Bigr)= \frac{e^{q_{s}d_{s}}}{2} \Biggl(\begin{matrix} 1 & -\frac{1}{\Lambda_{s} q_{s}} \\ -\Lambda_{s} q_{s}  & 1 \end{matrix} \Biggr) \Biggl(\begin{matrix} 1 & -\frac{1}{G} \\ 0 & 1 \end{matrix} \Biggr) \\ \nonumber
\Biggl(\begin{matrix} 1 & -\frac{d}{\Lambda_{f}} \\ -\Lambda_{f} q_{f}^{2} d & 1 \end{matrix} \Biggr) \Bigl(\begin{smallmatrix} T_{top}\\ F_{top} \end{smallmatrix} \Bigr)
\end{eqnarray}

\noindent where the substrate, interface conductance $G$, and the superlattice (replaced with a film) are each represented by a matrix. The film is optically thin and its approximately uniform absorption ($\alpha d \ll 1$) can be replaced by surface absorption ($\alpha d \gg 1$, where $\alpha$ is the absorption coefficient) as we do not consider processes on the $\frac{d^{2}}{D}\approx \frac{(10~nm)^{2}}{0.1~cm^{2}/s} = 10~ps$ timescale it takes heat to diffuse through the film. For simplicity, it is assumed first that the thermal properties of the glass near the metal film remain similar to those of the substrate. The condition $F_{b}=0$ gives (a term $ \frac{\Lambda_{s}q_{s}d}{\Lambda_{f}} \ll 1$ in the denumerator can be neglected)

\begin{eqnarray}
T_{top}(\omega,k)= F_{top}(\omega, k) \frac{1+\frac{\Lambda_{s}q_{s}}{G}}{\Lambda_{s}q_{s}+\Big (1+\frac{\Lambda_{s}q_{s}}{G} \Big) \Lambda_{f}q_{f}^{2}d}
\end{eqnarray}

\begin{eqnarray}
    =
\begin{cases}
    \frac{F_{top}}{\Lambda_{f}q_{f}^{2}d+\Lambda_{s}q_{s}} ,& \text{if }  \Lambda_{s}q_{s}\ll G  \text{~~~~~~~~~~(A)} \\
    \frac{F_{top}}{\Lambda_{f}q_{f}^{2}d + G },  & \text{if }  \Lambda_{s}q_{s}\gg G  \text{~~~~~~~~~~(B)}
\end{cases}
\end{eqnarray}

\noindent where two specific cases have been emphasized (figure 4). The incident heat flows along different paths, depending on the relation between the interface conductance $G$ and the substrate $\Lambda_{s}q_{s}$.

In case $A$ with no significant interface backscattering ($G\rightarrow \infty$ and identity interface matrix), cooling rates are limited by the substrate. The term $\Lambda_{f}q_{f}^{2} T_{top} d =\Lambda_{f}(k^{2}+\frac{i\omega C}{\Lambda_{f}}) T_{top} d $ represents the fraction of the incident flux that is carried away sideways in the film $\Lambda_{f} k^{2} T_{top} d$ or heats the film $i\omega C T_{top} d$ (figure 4(a)). Neglecting this term gives $T_{top}=\frac{F_{top}}{\Lambda_{s}q_{s}}$, the solution for a semi-infinite substrate with surface absorption, with a frequency dependence $T_{top}\propto \frac{1}{\sqrt{f}}$.

In case $B$, the interface conductance $G$ is low, limiting the cooling rate into the substrate. Then, $T_{top}=\frac{F_{top}}{\Lambda_{f} k^{2} d +\frac{i\omega \Lambda_{f}d}{D}+G}$. As above, the first two terms in the denominator represents sideways film flux and stored heat variations.

Unlike $\Lambda_{f} k^{2}d$, the new term $G$ remains finite as $k\rightarrow 0$ toward the peak of the pump profile, always giving a heat flow and removing the temperature divergence of the two-dimensional film in the steady-state (section 4.3). This term can be absorbed into the $\frac{i\omega \Lambda_{f}d}{D}$ factor, giving $\omega$ an imaginary part $i\frac{DG}{\Lambda_{f} d}$. This becomes an exponential $e^{-t/\tau_{G}}$ with the time constant $\tau_{G}=\frac{\Lambda_{f} d}{GD}=\frac{Cd}{G}$ following a Fourier transform, describing in the time domain the additional heat transfer channel opened through the interface.

The results can be slightly modified by replacing $F_{top}$ with the initial temperature $T_{initial}=\frac{F_{top}}{Cd}$. Then, the expressions in equation (4) relate $T_{final}$ and $T_{initial}$ and are the Green's functions in $\omega,k$ variables for different experimental conditions. For instance, $\mathbb{G}_{3D}(\omega,k)=\frac{1}{\Lambda_{s}q_{s}}$ (obtained in case $A$ in the limit $d \rightarrow 0$) is the three-dimensional Green's function for the substrate~\cite{2004Cahill}. Similarly $\mathbb{G}_{2D}(\omega,k)=\frac{1}{Dq_{f}^{2}}=\frac{1}{Dk^{2}+i\omega}$ (case $B$ in the limit $G\rightarrow \infty$) is the two-dimensional Green's function for the film. This is confirmed by a Hankel in $k$ and Fourier in $\omega$ transform of $\mathbb{G}_{2D}(\omega,k)$ that gives

\begin{eqnarray}
\mathbb{G}_{2D}(t,r)=\frac{\sqrt{2\pi}}{4tD}e^{-\frac{r^{2}}{4tD}}
\end{eqnarray}

\noindent the 2D Green's function in $t,r$ variables.

The Green's function in case $B$ can be written in a different useful form. The sideways heat flow can be replaced as $\Lambda k^{2}d \rightarrow \Lambda \frac{8d}{w_{0}^{2}}\rightarrow \frac{Cd}{\tau_{D}}$, representing the in-plane diffusion in the metal film with a characteristic time $\tau_{D}=\frac{w_{0}^{2}}{8D}$ in the time domain. Then, $\mathbb{G}=\frac{1}{\frac{1}{\tau_{D}}+i\omega+\frac{1}{\tau_{G}}}=\frac{1}{\frac{1}{\tau_{eff}}+i\omega}$, where $\frac{1}{\tau_{eff}}=\frac{1}{\tau_{D}}+\frac{1}{\tau_{G}}$ is the total rate at which heat leaves the layer, either through in-plane heat diffusion or through the interface. This expression for $\mathbb{G}$ shows that we can expect the phase of $T_{top}(\omega)$ to change on transitions from one interface-, film- or substrate-dominated cooling regime to another.

These results are applied to the two experimental configurations  -- the chopper modulation and the scanning beam.

\begin{figure}
\centering\rotatebox{0}{\includegraphics[scale=0.3]{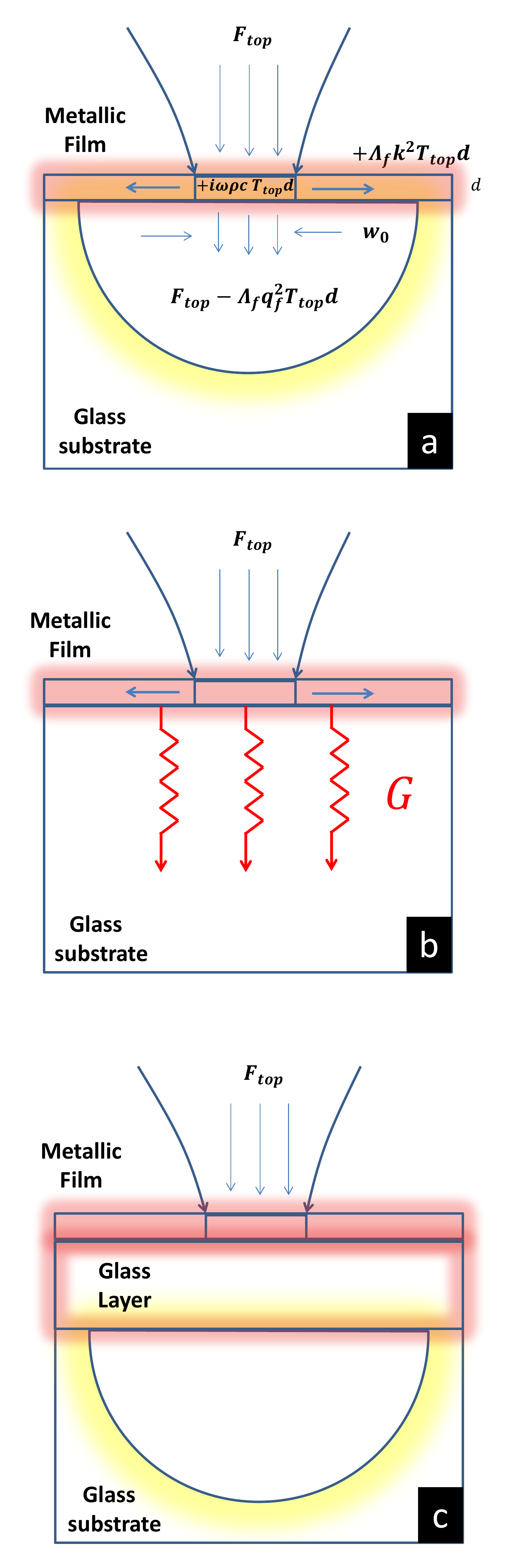}}
\caption{\label{fig:Figure1} (a) Conducting interface (case $A$) with cooling rates limited by the substrate. (b) A resistive interface (case $B$) limits the cooling rate. (c) Two-layer model, with a metallic and a glass film.}
\end{figure}

\subsection{Temporal modulation}

\begin{figure}
\centering\rotatebox{0}{\includegraphics[scale=0.4]{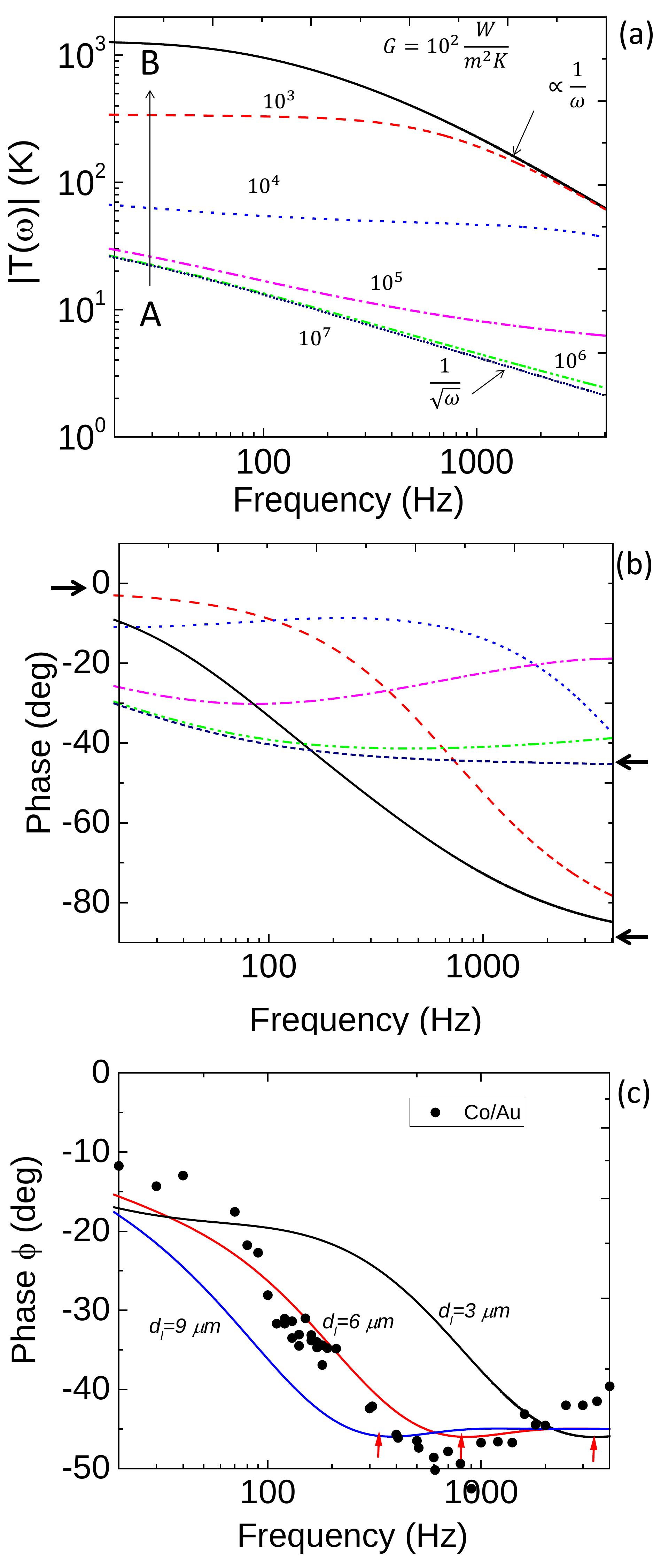}}
\caption{\label{fig:Figure1} Amplitude (a) and phase (b) of the temperature modulation for different $G$. As expected, the amplitude $|T|$ increases from $A$ to $B$. The parameters are $P_{abs}=30~mW$, $d = 30~nm$, $w_{eff} = 200~\mu m$, $D_{s} = 5 \times10^{-7}~m^{2}/s$, $\Lambda_{s} = 1~W/mK$, $D_{f} = 0.1\times10^{-4}~m^{2}/s$, $\Lambda_{f} = 100~W/mK$. (c) Results with a glass layer $d_{l}$ with $D_{l}=0.05D_{s}, \Lambda_{l}=0.05\Lambda_{s},w_{eff}=200~\mu m$. Similar results are obtained for layers with the same $\frac{D_{l}}{d_{l}^{2}}$ ratio.}
\end{figure}

In the pump-probe measurements the signal is proportional to the film temperature (equation (1)), which can be calculated with equation (4).

One may expect to observe heat diffusion following case $A$ because interfaces between dense solids have $G \gg \Lambda_{s}q_{s} \sim 10^{5}(1+i)~\frac{W}{m^{2}K}$~\cite{2014Cahill} for typical $k_{eff}$ and chopper frequencies $f$, or $T_{top}\propto \frac{1}{\sqrt{f}}$. Surprisingly, measurements show the $1/f$ dependence of two-dimensional heat flow above $200~Hz$ (figure 2) for Co/Pd and Co/Ag and a dependence between $1/f$ and $1/\sqrt{f}$ for Co/Au.

The $k-$ integral of equation (1) has been calculated numerically for different interface conductances $G$. The evolution of $T(\omega)$ amplitude and phase from case $A$ to $B$ is shown in figure 5(a)-(b). The main features of these plots can be understood qualitatively. In the limit of low frequency (steady-state), $T(\omega)$ is real and $\phi \rightarrow 0$. At large $\omega$ the Green's function dependence on $k$ is negligible and the $k-$ integral reduces to $\int_{0} ^{\infty} dk k e^{-\frac{k^{2}}{8}(w_{0}^{2}+w_{1}^{2})} =\frac{4}{w_{eff}^{2}}$. In case $A$ we obtain $T(\omega) \propto \frac{1}{\sqrt{i\omega}}$ or a $1/\sqrt{f}$ dependence and a $-45$ degree phase at large $\omega$. Similarly, in case $B$ we obtain $T(\omega) \propto \frac{1}{i\omega}$ or a $1/f$ and $-90$ degree phase at large $\omega$ (arrows).

The values of $G$ required to obtain the observed $1/f$ dependence (figure 2) are $\leq 10^{2}~\frac{W}{m^{2}K}$, much lower than typical glass-metal conductances $>10^{7} ~\frac{W}{m^{2}K}$~\cite{2009Juve} and on the order of the heat transfer coefficient for near-field radiative heat transfer between glass and Au interfaces at a $10~nm$ separation~\cite{2009Shen}. Such free-standing metal films would heat up to very high temperatures ($|T(\omega \rightarrow 0)|>10^{3}~K$ in figure 5(a)) and be structurally unstable. The minimum in phase at $1000~Hz$ is also not obtained.

The simplest geometry of one metallic film is insufficient and considering a more complicated structure is necessary. A solution is to add a new layer $l$ between the metal film and glass substrate (figure 4(c)). This layer may be the same porous birefringent layer, made in the glass substrate in the immediate vicinity of the absorbing film, and observed in polarizing microscopy. A glass layer of thickness $d_{l}$ and conductance to the glass substrate $G_{2} \gg \Lambda_{s}q_{s}$ will add one matrix $\rm \Biggl(\begin{matrix} \rm{cosh}(q_{l}d_{l}) & -\frac{1}{\Lambda_{l}d_{l}} \rm{sinh}(q_{l}d_{l}) \\ -\Lambda_{l} q_{l} \rm{sinh}(q_{l}d_{l}) & \rm{cosh}(q_{l}d_{l}) \end{matrix} \Biggr)$ to equation (3), where the small thickness approximation $q_{l}d_{l} \ll 1$ has not been made.

The result for $T_{top}(\omega,k)$ is unwieldy, but numerical calculations can be made for different layer thickness $d_{l}$. These show a characteristic minimum (arrows in figure 5(c)), as the phase climbs to the semi-infinite substrate angle $-45$ degrees. This occurs near $f=\frac{D}{\pi d_{l}^{2}}$, when the depth of thermal modulation is smaller than the layer thickness and the sample begins to resemble a semi-infinite substrate. The minimum in phase observed at $1000~Hz$ is obtained. Measurements for the Co/Au sample from figure 2(b) have been shifted along the $y-$axis and added for comparison. The model explains the main experimental features.

\subsection{Demagnetization patterns}

\begin{figure}
\centering\rotatebox{0}{\includegraphics[scale=0.5]{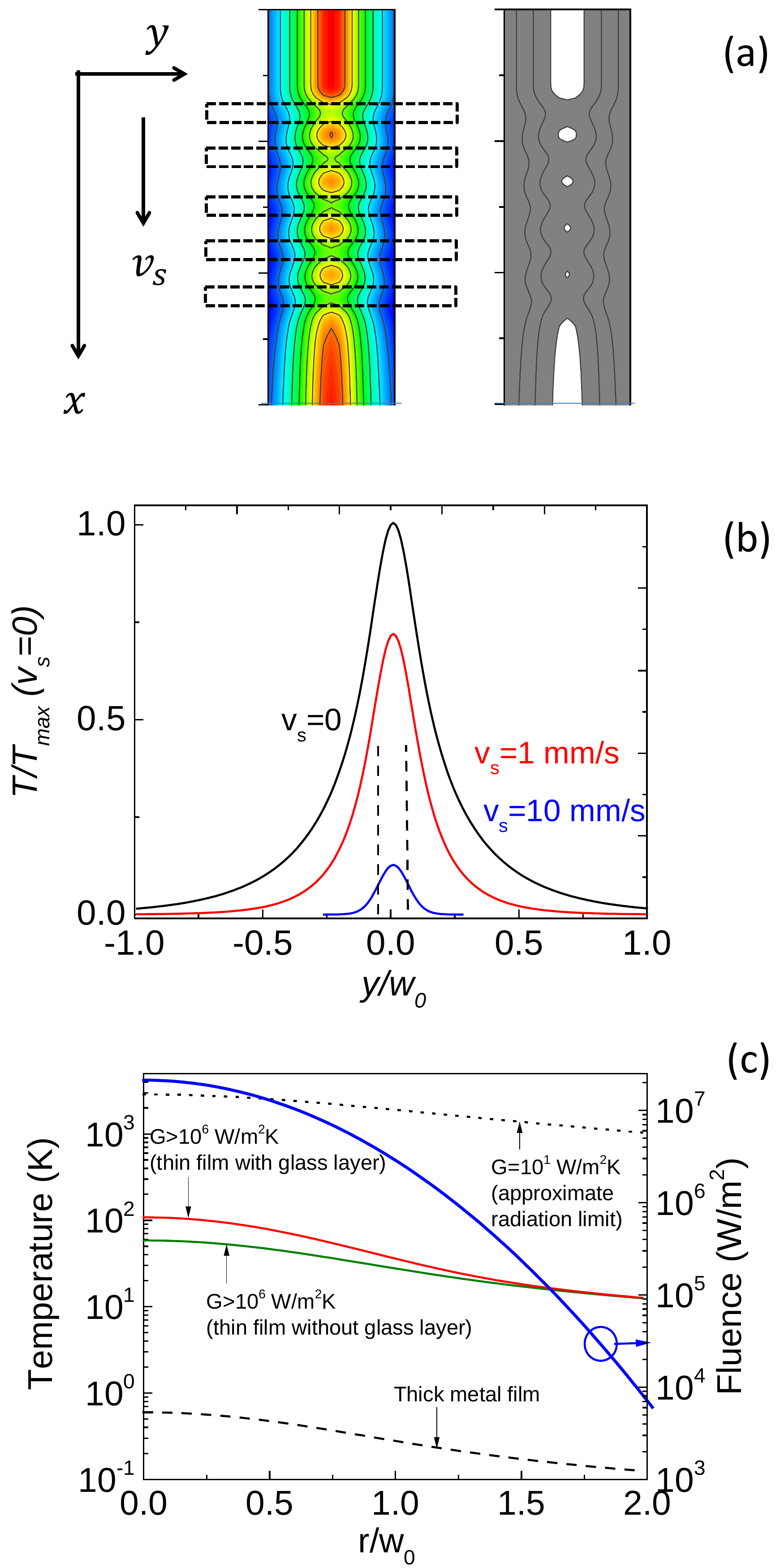}}
\caption{\label{fig:Figure1} (a) Film $T_{max}$ during a scan with power modulated across the horizontal stripes (dashed features). (b) The steady-state solution at $v_{s}=0$ compared to a profile induced by a beam moving with $v_{s}=1,10~mm/s$ shows the larger radius at the end of the scan. (c) Steady-state temperature radial temperature profiles for interface conductances $G$ shown and $P_{abs}=30~mW$, $d = 30~nm$, $w_{0} = 200~\mu m$, $D_{s} = 5\times 10^{-7}~m^{2}/s$, $\Lambda_{s} = 1~W/mK$, $D_{f} = 0.1 \times10^{-4}~m^{2}/s$, $\Lambda_{f} = 100~W/mK$ without the glass layer, compared to the 3D case (thick metal film), and with results including the same glass layer as in figure 5(c). The difference between results for $G=10^{6}~W/m^{2}K$ and $G\rightarrow \infty$ is negligible. The right axis shows the Gaussian beam fluence profile.}
\end{figure}

Only the pump beam is present in scanning measurements and its pulse sequence is replaced with a continuous-wave beam as before. In contrast to the previous case, the cylindrical symmetry is lost and the initial temperature spectrum is not sharply defined for a moving beam. It is advantageous to work with $t,r$ variables.

A Gaussian beam moving at a velocity $v_{s}$ along the $x-$ axis is a heat source $F(t',x',y')=\frac{2P_{abs}(t')}{\pi w_{0}^{2}}e^{-\frac{2(x'-v_{s}t')^{2}}{w_{0}^{2}}-\frac{2(y')^{2}}{w_{0}^{2}}}$ giving an initial temperature $T(t',x',y')=\frac{F(t',x',y')}{Cd}$, where the absorbed power depends on time $t'$ because of the variable transmission observed on crossing stripes (figure 2). The temperature at a later time $t>t'$ is $T(t,x,y)=\int dt'dx'dy' \mathbb{G}(t-t',x-x',y-y') T(t',x',y')$. The solution with the 3D Green's function for a substrate with surface absorption has been obtained before~\cite{1980Nissim}. For the 2D Green's function (equation 6) the integration over $x'$ and $y'$ can be done by completing the square to give

\begin{eqnarray}
T(t,x,y)= \frac{2\sqrt{2\pi}}{Cd} \int_{-\infty} ^{t} dt'\frac{P_{abs}(t')}{8D(t-t')+w_{0}^{2}} e^{-2\frac{(x-v_{s}t')^{2}+y^{2}}{8D(t-t')+w_{0}^{2}}}
\end{eqnarray}

\noindent This expression shows how in-plane diffusion (the $8D(t-t')$ term) combines with the beam profile tails (the $w_{0}^{2}$ term) to give a $T$ increase at $(x,y)$ when the laser beam center is at $(v_{s}t',0)$.

As expected, this integral is divergent for a stationary beam of constant intensity ($x=0,y=0,v_{s}=0, P_{abs}(t')=const.$). In contrast to a 3D substrate, a thermally 2D film does not cool well under a steady heat flux. To remove this unphysical divergence, the heat flow conditions must be changed from strictly two-dimensional. The Green's function for a layer with surface absorption and an infinite interface conductance $G\rightarrow \infty$ to the substrate can be reduced to a double integral~\cite{1982Burgener}. A different approach can be taken for an interface with a finite conductance $G$, by allowing an additional heat transfer channel through the interface, as done in section 4.1, or

\begin{eqnarray}
\mathbb{G}=\mathbb{G}_{2D}(t-t',x-x',y-y') e^{-\frac{t-t'}{\tau_{G}}}
\end{eqnarray}

This removes the temperature divergence. Results of the maximum film temperature $T_{max}(x,y)$ obtained during the laser scan for a variable $P_{abs}$, modulated as shown in figure 2 give a sequence of peaks (figure 6(a), left panel). Dividing into two  cases (white and black, figure 6(a), right panel), above and below a borderline temperature gives plots that correspond well with the experimental observations (figure 3). In particular, the decreasing spot size and re-emergence of the stripe on crossing the stack is obtained. A narrowing of the features, calculated for a moving beam (figure 6(b)), is consistent with observations (figure 3(c)).

The cylindrical symmetry is restored in the stationary condition ($v_{s}=0$) and this case can be applied to illustrate the temperature increase due to heat accumulation. In the substrate limit (case $A$ with $d\rightarrow 0$) a Hankel transform of $T_{top}(\omega=0,k)=\frac{F_{top}(\omega=0,k)}{q_{s}\Lambda_{s}}$ gives the known solution $T_{top}(\omega=0,r)=\frac{1}{\sqrt{2 \pi}}\frac{P_{abs}}{\Lambda_{s} w_{0}} e^{-\frac{r^{2}}{w_{0}^{2}}}I_{0}\Big(\frac{r^{2}}{w_{0}^{2}}\Big)$, where $I_{0}$ is the modified Bessel function~\cite{2011Bauerle}. Then $T_{max}=\frac{P_{abs}}{\sqrt{2 \pi}\Lambda w_{0}} \approx \frac{30~mW}{2.5\times 100~W/mK ~200~\mu m}\sim 0.6~K $ for typical metallic thermal conductivity $\Lambda$. Heat dissipates quickly between pulses and $T_{acc}$ can be neglected in thick, thermally conducting, samples.

In contrast, our superlattice samples are thermally thin. In general, an interface at a depth smaller than the thermal modulation depth $L_{th}=\sqrt{\frac{D}{\pi f}}$ will affect heat diffusion, where $L_{th}=30~\mu m \gg d$ for $f=3\times10^{3}~Hz$ and a good thermal conductor with $D=0.1 \times10^{-4}~\frac{m^{2}}{s}$. Temperature profiles calculated with a Hankel transform of equation (4) are significantly higher as the large heat fluxes possible through a semi-infinite substrate are reduced (figure 6(c)). The combination of large light absorption in the metallic film and small glass thermal conductivity can raise $T_{acc}$ above that of a thick metallic film. Heat accumulation and large temperature gradients in our samples explain the observed forces on magnetic domain walls during all-optical switching~\cite{2017F} and how a final demagnetized state can be obtained from both heat accumulation $T_{acc}$ and transient $T_{tr}$ ultrafast demagnetization~\cite{2017F-arXiv}.

A large $T_{acc}$ in magnetic materials can be beneficial (for instance, in heat-assisted magnetic recording) or undesired if measurements at low temperature are required. Future work may examine the smaller heat accumulation predicted in one-dimensional structures at the same average fluence with increased repetition rates, heat accumulation in dots, offset pump-probe beams or conditions with a larger $k_{eff}$ from tighter focusing.

\section{Conclusion}

Heat diffusion in metallic superlattices on glass substrates has been examined with pump-probe measurements and polarizing microscopy of laser-induced demagnetization patterns. Green's function solutions of the heat diffusion equation quantify the temperature in the two experimental configurations. A glass layer is required to explain the temporal modulation frequency dependence and demagnetization patterns are reproduced with an interface conductance. Thermo-transmittance measurements can be applied in examining heat accumulation and diffusion in thin samples on thermally insulating substrates under an intense light field and in characterization of a multilayer device thermal response.

\section*{Acknowledgments}

The authors would like to thank X Wang and J Aebersold at the University of Louisville cleanroom for assistance. This research was supported by the University of Louisville Research Foundation.

\begin{figure}
\centering\rotatebox{0}{\includegraphics[scale=0.3]{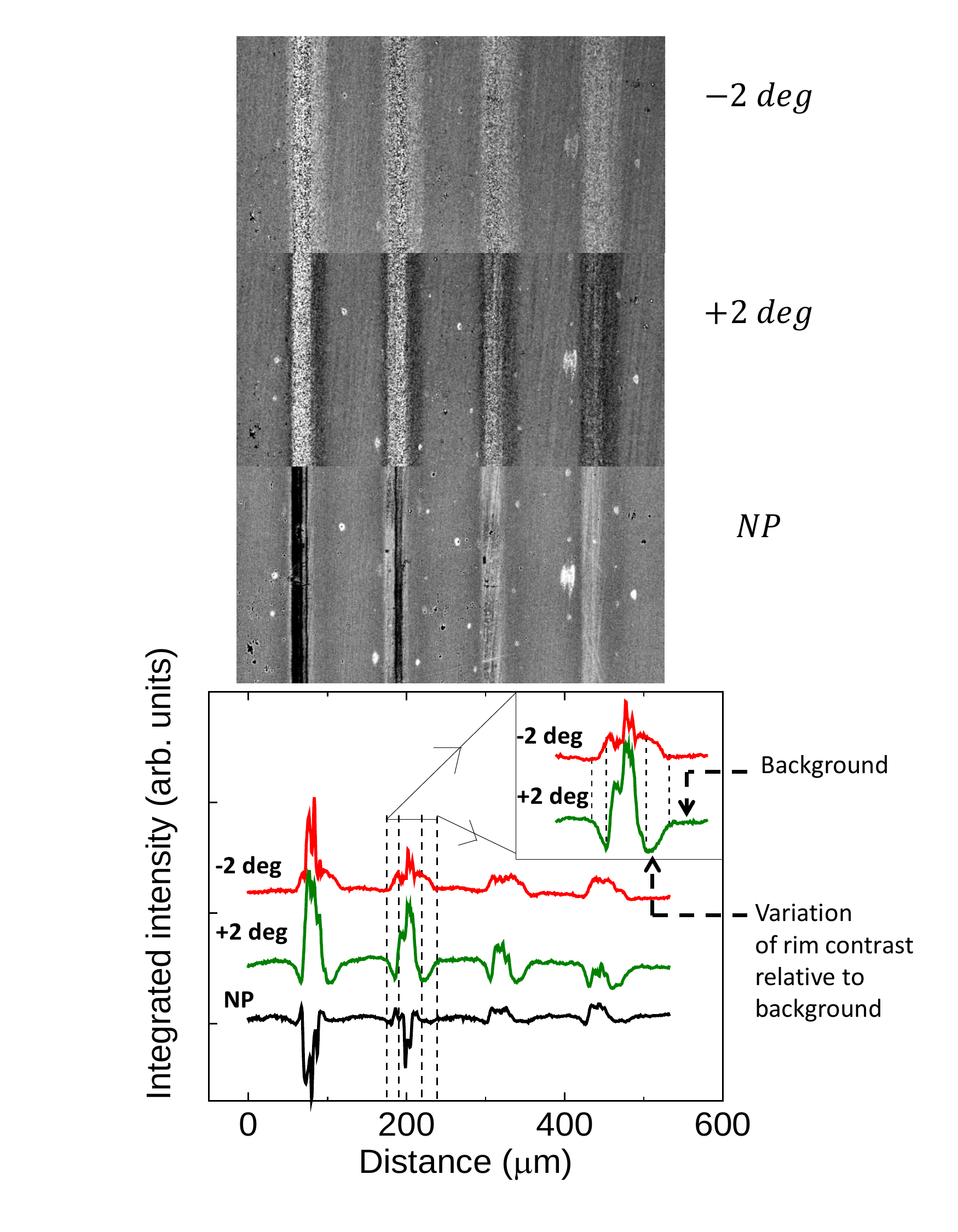}}
\caption{\label{fig:Figure1} Supplementary figure 1. Top panel: polarizing microscopy images with the polarizer and analyzer removed (labelled ``NP"), and with ``$-2$ deg" and ``$+2$ deg" relative rotations from extinction condition. Birefringent areas show a variation in contrast with respect to the background. Lower panel: intensity averaged along the vertical direction. A difference between the left and right edges is due to a slightly asymmetric beam profile.}
\end{figure}


\section*{References}

\begin{thebibliography}{}

\bibitem{2007Fretigny}
Ch. Fretigny, J. P. Roger, V. Reita and D. Fournier \emph{Analytical inversion of photothermal measurements: Independent determination of the thermal conductivity and diffusivity of a conductive layer deposited on an insulating substrate}, J. Appl. Phys. \textbf{102}, 116104 (2007)

\bibitem{1994Reichling}
M. Reichling and H. Gronbeck \emph{Harmonic heat flow in isotropic layered systems and its use for thin film thermal conductivity measurements}, J. Appl. Phys. \textbf{75 (4)}, 1914 (1994)

\bibitem{2014Cahill}
D. G. Cahill, P. V. Braun, G. Chen, D. R. Clarke, S. Fan, K. E. Goodson, P. Keblinski, W. P. King, G. D. Mahan, A. Majumdar, H. J. Maris, S. R. Phillpot, E. Pop E L. and Shi \emph{Nanoscale thermal transport. II. 2003 -- 2012}, Appl. Phys. Rev. \textbf{1}, 011305 (2014)

\bibitem{2008Schmidt-a}
A. Schmidt, M. Chiesa, X. Chen and G. Chen \emph{An optical pump-probe technique for measuring the thermal conductivity of liquids}, Rev. Sci. Instr. \textbf{79}, 064902 (2008)

\bibitem{2008Schmidt-b}
A. Schmidt, X. Chen and G. Chen \emph{Pulse accumulation, radial heat conduction, and anisotropic thermal conductivity in pump-probe transient thermoreflectance}, Rev. Sci. Instr. \textbf{79}, 114902 (2008)

\bibitem{2017deJong}
J. A. de Jong, A. M. Kalashnikova, R. V. Pisarev, A. M. Balbashov, A. V. Kimel, A. Kirilyuk and Th. Rasing \emph{Effect of laser pulse propagation on ultrafast magnetization dynamics in a birefringent medium}, J. Phys.: Condens. Matter \textbf{29}, 164004 (2017)

\bibitem{2017Afanasiev}
D. Afanasiev, B. A. Ivanov, R. V. Pisarev, A. Kirilyuk, Th. Rasing and A. V. Kimel \emph{Femtosecond single-shot imaging and control of a laser-induced first-order phase transition in $\rm HoFeO_3$}, J. Phys.: Condens. Matter \textbf{29}, 224003 (2017)

\bibitem{2007Stanciu}
C. D. Stanciu, F. Hansteen, A. V. Kimel, A. Kirilyuk, A. Tsukamoto, A. Itoh and Th. Rasing \emph{All-Optical Magnetic Recording with Circularly Polarized Light}, Phys. Rev. Lett. \textbf{99}, 047601 (2007)

\bibitem{2006Aktag}
A. Aktag, S. Michalski, L. Yue, R. D. Kirby and S-H. Liou \emph{Formation of an anisotropy lattice in Co/Pt multilayers by direct laser interference patterning}, J. Appl. Phys. \textbf{99}, 093901 (2006)

\bibitem{2006Schuppler}
C. Schuppler, A. Habenicht, I. L. Guhr, M. Maret, P. Leiderer, J. Boneberg and M. Albrecht \emph{Control of magnetic anisotropy and magnetic patterning of perpendicular Co/Pt multilayers by laser irradiation}, Appl. Phys. Lett. \textbf{88}, 012506 (2006)

\bibitem{2009Leufke}
P. M. Leufke, S. Riedel, M-S. Lee, J. Li, H. Rohrmann, T. Eimuller, P. Leiderer, J. Boneberg, G. Schatz and M. Albrecht \emph{Two different coercivity lattices in Co/Pd multilayers generated by single-pulse direct laser interference lithography}, J. Appl. Phys. \textbf{105}, 113915 (2009)

\bibitem{2014Kisielewski}
J. Kisielewski, W. Dobrogowski, Z. Kurant, A. Stupakiewicz, M. Tekielak, A. Kirilyuk, A. Kimel , Th. Rasing, L. T. Baczewski, A. Wawro, K. Balin, J. Szade and A. Maziewski \emph{Irreversible modification of magnetic properties of Pt/Co/Pt ultrathin films by femtosecond laser pulses}, J. Appl. Phys. \textbf{115}, 053906 (2014)

\bibitem{2015Stark}
M. Stark, F. Schlickeiser, D. Nissen, B. Hebler, P. Graus, D. Hinzke, E. Scheer, P. Leiderer, M. Fonin, M. Albrecht, U. Nowak and J. Boneberg \emph{Controlling the magnetic structure of Co/Pd thin films by direct laser interference patterning}, Nanotechnology \textbf{26}, 205302 (2015)

\bibitem{2014Lambert}
C-H. Lambert, S. Mangin, B. Varaprasad, Y. K. Takahashi, M. Hehn, M. Cinchetti, G. Malinowski, K. Hono, Y. Fainman, M. Aeschlimann and E. E. Fullerton \emph{All-optical control of ferromagnetic thin films and nanostructures}, Science \textbf{345}, 1337 (2014)

\bibitem{2014Mangin}
S. Mangin, M. Gottwald, C-H. Lambert, D. Steil, V. Uhlír, L. Pang, M. Hehn, S. Alebrand, M. Cinchetti, G. Malinowski, Y. Fainman, M. Aeschlimann and E. E. Fullerton \emph{Engineered materials for all-optical helicity-dependent magnetic switching} Nat. Materials \textbf{13}, 286 (2014)

\bibitem{2017Gorchon}
J. Gorchon, C-H. Lambert, Y. Yang, A. Pattabi, R. B. Wilson, S. Salahuddin and J. Bokor \emph{Single shot ultrafast all optical magnetization switching of ferromagnetic Co/Pt multilayers}, Appl. Phys. Lett. \textbf{111}, 042401 (2017)

\bibitem{2016Takahashi}
Y. K. Takahashi, R. Medapalli, S. Kasai, J. Wang, K. Ishioka, S. H. Wee, O. Hellwig, K. Hono and E. E. Fullerton \emph{Accumulative Magnetic Switching of Ultrahigh-Density Recording Media by Circularly Polarized Light}, Phys. Rev. Applied \textbf{6}, 054004 (2016)

\bibitem{2017John}
R. John, M. Berritta, D. Hinzke, C. Muller, T. Santos, H. Ulrichs, P. Nieves, J. Walowski, R. Mondal, O. Chubykalo-Fesenko, J. McCord, P. M. Oppeneer, U. Nowak and M. Munzenberg \emph{Magnetisation switching of FePt nanoparticle recording medium by femtosecond laser pulses} Sci. Reports \textbf{7}, 4114 (2017)

\bibitem{2016Cornelissen}
T. D. Cornelissen, R. Cordoba and B. Koopmans \emph{Microscopic model for all optical switching in ferromagnets} Appl. Phys. Lett. \textbf{108}, 142405 (2016)

\bibitem{2012Khorsand}
A. R. Khorsand, M. Savoini, A. Kirilyuk, A. V. Kimel, A. Tsukamoto, A. Itoh and Th. Rasing \emph{Role of Magnetic Circular Dichroism in All-Optical Magnetic Recording}, Phys. Rev. Lett. \textbf{108}, 127205 (2012)

\bibitem{2012Ostler}
T. A. Ostler, J. Barker, R. F. L. Evans, R. W. Chantrell, U. Atxitia, O. Chubykalo-Fesenko, S. El Moussaoui, L. Le Guyader, E. Mengotti, L. J. Heyderman, F. Nolting, A. Tsukamoto, A. Itoh, D. Afanasiev, B. A. Ivanov, A. M. Kalashnikova, K. Vahaplar, J. Mentink, A. Kirilyuk, Th. Rasing and A. V. Kimel \emph{Ultrafast heating as a sufficient stimulus for magnetization reversal in a ferrimagnet}, Nat. Comm. \textbf{3}, 666 (2012)

\bibitem{2016Xu}
C. Xu, T. A. Ostler and R. W. Chantrell \emph{Thermally induced magnetization switching in Gd/Fe multilayers}, Phys. Rev. B \textbf{93}, 054302 (2016)

\bibitem{2016Ellis}
M. O. A. Ellis, E. E. Fullerton and R. W. Chantrell \emph{All-optical switching in granular ferromagnets caused by magnetic circular dichroism}, Sci. Rep. 6, 30522 (2016)

\bibitem{2016Gorchon}
J. Gorchon, Y. Yang and J. Bokor \emph{Model for multishot all-thermal all-optical switching in ferromagnets}, Phys. Rev. B 94, 020409(R) (2016)

\bibitem{2016Hadri}
M. S. El Hadri, P. Pirro, C-H. Lambert, S. Petit-Watelot, Y. Quessab, M. Hehn, F. Montaigne, G. Malinowski and S. Mangin \emph{Two types of all-optical magnetization switching mechanisms using femtosecond laser pulses}, Phys Rev B \textbf{94}, 064412 (2016)

\bibitem{2017F}
F. Hoveyda, E. Hohenstein and S. Smadici \emph{Heat accumulation and all-optical switching by domain wall motion in Co/Pd superlattices}, J. Phys.: Condens. Matter \textbf{29}, 225801 (2017)

\bibitem{2004Cahill}
D. G. Cahill \emph{Analysis of heat flow in layered structures for time-domain thermoreflectance}, Rev. Sci. Instr. \textbf{75(12)}, 5119 (2004)

\bibitem{2009Schmidt}
A. J. Schmidt, R. Cheaito and M. Chiesa \emph{A frequency-domain thermoreflectance method for the characterization of thermal properties} Rev. Sci. Instr. \textbf{80}, 094901 (2009)

\bibitem{2014Liu}
J. Liu, G-M. Choi and D. G. Cahill \emph{Measurement of the anisotropic thermal conductivity of molybdenum disulfide by the time-resolved magneto-optic Kerr effect}, J. Appl. Phys. \textbf{116}, 233107 (2014)

\bibitem{2000Qiu}
Z. Q. Qiu and S. D. Bader \emph{Surface magneto-optic Kerr effect}, Rev. Sci. Instr. \textbf{71(3)}, 1243 (2000)

\bibitem{2017F-arXiv}
F. Hoveyda, E. Hohenstein, R. Judge and S. Smadici, to be submitted.

\bibitem{2010Wang}
Y. Wang, J-Y. Park, Y-K. Koh and D. G. Cahill \emph{Thermoreflectance of metal transducers for time-domain thermoreflectance}, J. Appl. Phys. \textbf{108}, 043507 (2010)

\bibitem{2012Wilson}
R. B. Wilson, B. A. Apgar, L. W. Martin and D. G. Cahill \emph{Thermoreflectance of metal transducers for optical pump-probe studies of thermal properties}, Opt. Express \textbf{20(27)}, 28829 (2012)

\bibitem{1993Jewell}
J. M. Jewell \emph{Thermooptic Coefficients of Soda-Lime-Silica Glasses}, J. Am. Ceram. Soc. \textbf{76(7)} 1855 (1993)

\bibitem{2001Sudrie}
L. Sudrie, M. Franco, B. Prade and A. Mysyrowicz \emph{Sudy of damage in fused silica induced by ultra-short IR laser pulses} Opt. Comm. \textbf{191}, 333 (2001)

\bibitem{2001Schaffer}
C. B. Schaffer, A. Brodeur, J. F. García and E. Mazur \emph{Micromachining bulk glass by use of femtosecond laser pulses with nanojoule energy}, Opt. Lett. \textbf{26(2)}, 93 (2001)

\bibitem{2003Schaffer}
C. B. Schaffer, J. F. García and E. Mazur \emph{Bulk heating of transparent materials using a high-repetition-rate femtosecond laser}, Appl. Phys. A \textbf{76}, 351 (2003)

\bibitem{2005Hnatovsky}
C. Hnatovsky, R. S. Taylor, E. Simova, V. R. Bhardwaj, D. M. Rayner and P. B. Corkum \emph{Polarization-selective etching in femtosecond laser-assisted microfluidic channel fabrication in fused silica}, Opt. Lett. \textbf{30(14)}, 1867 (2005)

\bibitem{2009Cheng}
G. Cheng, K. Mishchik, C. Mauclair, E. Audouard and R. Stoian \emph{Ultrafast laser photoinscription of polarization sensitive devices in bulk silica glass}, Opt. Express \textbf{17(12)}, 9515 (2009)

\bibitem{2009Juve}
V. Juve, M. Scardamaglia, P. Maioli, A. Crut, S. Merabia, L. Joly, N. Del Fatti and F. Vallee \emph{Cooling dynamics and thermal interface resistance of glass-embedded metal nanoparticles}, Phys. Rev. B \textbf{80}, 195406 (2009)

\bibitem{2009Shen}
S. Shen, A. Narayanaswamy and G. Chen \emph{Surface Phonon Polaritons Mediated Energy Transfer between Nanoscale Gaps}, Nano Letters \textbf{9(8)}, 2909 (2009)

\bibitem{1980Nissim}
Y. I. Nissim, A. Lietoila, R. B. Gold and J. F. Gibbons \emph{Temperature distributions produced in semiconductors by a scanning elliptical or circular cw laser beam}, J. Appl. Phys. \textbf{51(1)}, 274 (1980)

\bibitem{1982Burgener}
M. L. Burgener and R. E. Reedy \emph{Temperature distributions produced in a two-layer structure by a scanning cw laser or electron beam}, J. Appl. Phys. \textbf{53(6)}, 4357 (1982)

\bibitem{2011Bauerle}
D. Bauerle \emph{Laser processing and chemistry} 4th edition, Springer Verlag (2011)

\end{thebibliography}
\end{document}